\documentclass[showpacs,preprintnumbers,amsmath,amssymb,preprint]{revtex4}
\usepackage{graphics}
\begin{document}

\title{
Isospin dependent thermodynamics of fragmentation}

\author{Ad. R. Raduta$^{(1,2)}$} 
\author{F. Gulminelli$^{(3)}$} 
\thanks{member of the Institut Universitaire de France}
\affiliation{
	$^{1}$~NIPNE, Bucharest-Magurele, POB-MG 6, Romania  
	\\
	$^{2}$~Institut de Physique Nucleaire, 
	IN2P3-CNRS, F-91406 Orsay c\'{e}dex, France\\
	$^{3}$~LPC (IN2P3-CNRS/Ensicaen et Universit\'{e}), 
	F-14050 Caen c\'{e}dex, France}

\begin{abstract}
The thermal and phase properties of a multifragmentation model 
which uses clusters as degrees of freedom, are explored as a function of
isospin. A good qualitative agreement is found with the phase diagram of
asymmetric nuclear matter as established by different mean-field models.
In particular, from the convexity properties of the nuclear entropy, 
we show that uncharged finite nuclei display first and second order 
liquid-gas-like phase transitions. 
Different quantities are examined to connect the thermal properties 
of the system to cluster observables.
In particular we show that fractionation is only a loose 
indication of phase coexistence.
A simple analytical formula is proposed and tested to evaluate the 
symmetry (free) energy from the widths of isotopic distributions.
Assuming that one may restore the isotopic composition of break-up fragments,
it is found that some selected isotopic observables can allow 
to quantitatively access the freeze-out symmetry energy in
multifragmentation experiments.

\end{abstract}
\date{\today}
\maketitle

\section{Introduction} 
\label{intro}

Due to the short range repulsive and finite range attractive character of the
nucleon-nucleon interaction, nuclear matter
is known to exhibit a phase transition similar to the liquid-gas
transition taking place in real fluids \cite{Finn,Bertsch}.

Many experimental and theoretical efforts have been devoted to 
this subject \cite{Das-PhysRep}; in particular,
most of the recent works focus on asymmetric matter and the
effect of isospin as an additional degree of freedom
\cite{muellerserot,koonin,leemekjian,sil}. 
Mean-field based models have demonstrated the presence 
of a first order phase transition in both isospin symmetric 
and asymmetric nuclear matter \cite{gulminelli}, 
the decrease of the critical temperature with increasing 
isospin asymmetry and, in the case of asymmetric matter, 
different neutron/proton compositions of the liquid and gas phases
\cite{muellerserot,leemekjian,sil,gulminelli}. 

In astrophysics, both neutron-star structure and supernova dynamics 
are influenced by thermal properties of neutron-rich nuclear matter 
in a large interval of temperatures and densities 
\cite{Lattimer-PhysRep, Glendenning}. 
All these different phenomena involve excited matter at baryon 
densities lower than normal nuclear 
matter density: this corresponds in the phase diagram to a region of 
instability with respect to phase separation. Information on the 
phase structure and properties of hot and diluted nuclear matter
is thus of clear astrophysical relevance.

From an experimental point of view, the only terrestrial phenomenon
that may allow to access finite temperature low density properties 
of neutron rich matter, and in particular pin down the density and temperature
dependence of the symmetry energy, 
is given by nuclear multifragmentation \cite{Das-PhysRep}.
Indeed, this specific decay channel of nuclei, whose excitation energy is of
the order of a few MeV/nucleon, has been since a long time 
tentatively associated to the coexistence zone of the nuclear 
matter phase diagram \cite{richert-PhysRep}.

However, the connection between multifragmentation observables 
and the nuclear matter energy-density functional and phase structure is 
far from being trivial. First, finite nuclei are charged while 
nuclear matter is by definition neutral. 
Therefore, the presence of this non-saturating long range force makes 
difficult to rely the multifragmentation phenomenology to the 
theoretical studies of nuclear matter
\footnote{It may be interesting to remark that, on the other hand, 
the presence of the Coulomb interaction may make finite nuclei 
a good laboratory to address the properties 
of compact stellar objects. Indeed, neutrality is verified only on a 
macroscopic scale in neutron star crust matter, and charge fluctuations are 
recognized to be at the origin of the crust phase structure \cite{watanabe}.}.
Second and even more important, atomic nuclei are composed of
a very small number of constituents, 
and their phase properties are not trivially 
linked to the phase structure of nuclear matter. 
At the thermodynamic limit, the coexistence zone is a simple linear 
superposition of pure liquid and gas phases,
and can be deduced from a mean field approach through a Gibbs construction.
The situation is completely different in finite nuclei, where
phase coexistence is revealed by convexity anomalies of the entropy surface
\cite{gross,cg-phd}. The properties of coexistence cannot 
be deduced from the properties of the pure phases (nuclei and nucleons, 
respectively), mean-field based approaches badly fail, and a description 
explicitly accounting  for complex clusterization is mandatory \cite{gross}.

A typical example of this ambiguity is given by the fractionation phenomenon,
originally expounded as a consequence of Gibbs phase equilibrium,
later on recognized as a generic feature of
cluster formation \cite{baoanli,ono,smf}, and systematically observed in the 
experimental analyses \cite{xu,geraci,martin,shetty,botvina}:
a clear thermodynamic interpretation of fractionation is not trivial in 
a finite system
since we cannot unambiguously associate a given fragment size to the "liquid" 
or "gas" phase. 
 
To contribute to match the gap between nuclear matter thermodynamic studies 
and nuclear multifragmentation, in this paper we investigate the isospin 
dependent phase diagram of finite excited nuclei.
The study is done in the framework of the 
Microcanonical Multifragmentation Model (MMM) \cite{mmm}, 
that explicitly considers clusters as degrees of freedom.
A similar analysis was already presented in Ref. \cite{npa2002},
where Coulomb effects on the phase diagram were especially addressed.
Here, in order to concentrate on isospin effects, 
we first consider that the Coulomb interaction is switched off.
The use of such an idealized neutral system will 
additionally allow us to make connections
with the expected behavior at the thermodynamic limit and
to explore the link between isospin observables 
(isoscaling \cite{ono,tsang,LeFevre,botvina,shetty04,Tsang2,Ono2},
isospin fluctuations \cite{botvina,xu,Liu,ColonnaMatera,Wen})
and the low density finite temperature symmetry (free) energy
of the equation of state.
The robustness of measurements of the symmetry energy in both
finite neutral systems and real nuclei under the effect of
mass and charge conservation and Coulomb is finally addressed.

\section{Theoretical framework}
\label{theory}

To provide a realistic description of multifragment production, 
a theory dealing with complex correlations well beyond 
the mean field is needed. 
An exact solution of this problem at the microscopic level is provided
by classical models (Molecular Dynamics, Lattice Gas) \cite{richert-PhysRep},
which however completely miss all the specific quantal features of the process.
Moreover, no connection is possible within these models between fragment 
observables and the different ingredients of the nuclear energy-density 
functional, which are explored through multifragmentation reactions.
On the other hand, semiclassical or
quantum molecular models (such as QMD \cite{qmd}, AMD \cite{amd}) which
reproduce the most important macroscopic nuclear properties as
density distributions and binding energies, and account for
nucleon-nucleon interaction and Pauli blocking, still do not offer
a definite description of multifragmentation 
and access to the equation of state of excited matter
because of the ambiguity of the fragments and break-up definition.

An interesting alternative is given by
statistical models which use clusters as degrees of freedom \cite{fisher}.
Such models offer the remarkable advantage that all nuclear bound as well as 
continuum states are naturally accounted for via empirical parameterizations
of the clusters energies and level densities, allowing a
direct comparison with experimental data.

The price to pay for such a realistic inclusion of nuclear effects,
is the underlying hypothesis that nuclear correlations are entirely
exhausted by clusterization, which amounts to the implementation 
of the properties of isolated low excited nuclei 
for the description of break-up fragments.
This limitation can in principle be avoided by including
effective in medium corrections in the fragments energy functional
\cite{mishustin}.
In this case, however, the fragment energy becomes a free parameter
of the theory.

The distinctive features of the fragmentation process, namely 
the explosion of an isolated nucleus into
vacuum, recommends the microcanonical framework \cite{mmmc,smm,randrup}
as the most natural choice.
The unphysical hypothesis of a sharp fixed freeze-out volume constraint
may be easily overcome by considering a total spatial extension for the 
fragmenting system fluctuating event by event \cite{annals}.
Technically, this is realized introducing a $\lambda$ Lagrange parameter conjugate
of the volume $V$ which 
alters the statistical weight 
of a configuration $W_C$ by an extra factor, $\exp(-\lambda V)$ 
\cite{constantbP}.
The thermodynamical potential associated to this ensemble
is $\bar{S}_E \left[ \lambda \right]=S- \lambda V=\ln {\cal W}(E,\lambda)$,
where ${\cal W}(E,\lambda)=\int W(E,V) \exp(-\lambda V) {\rm d}V$.
In addition, for a system belonging to the liquid-gas universality class, 
the exploration of the configuration space along constant $\lambda$
paths provides a straightforward method to reveal phase coexistence
by the back-bendings of the corresponding caloric curves and to
finally construct the phase diagram \cite{cg-phd}.

The MMM version \cite{mmm} of the microcanonical multifragmentation models
\cite{mmmc,smm,randrup}
has been used so far to investigate the thermodynamic properties of charged 
nuclei with excitation energies between 1 and 15 MeV/nucleon \cite{npa2002},
and will be presently employed for the study of isospin effects.
MMM provides a Monte-Carlo calculation of the global density of states 
$W(A,Z,E,{\rm P},{\rm L},V)$ of a nuclear
system modelized as a non-interacting collection of nuclear clusters.
The space of observables is given by the baryonic number $A$, 
the proton number $Z$, 
total energy $E$, total momentum ${\rm P}$, total angular momentum ${\rm L}$
and freeze-out volume $V$.
The investigation of all clusters states compatible with conservation laws and
geometrical restrictions is performed using a Metropolis trajectory in 
configuration space.

Break-up fragments are considered as having normal nuclear density $\rho_0$,
and described by a ground-state liquid drop binding energy including surface
and symmetry terms. This description is consistent with a 
semiclassical Thomas-Fermi approximation \cite{ringschuck} 
or hot Hartree-Fock \cite{vautherin}, 
where the effect of temperature is a modified occupation of the single 
particle eigenstates of the mean field hamiltonian. The finite temperature 
fragment energy functional in this approach is thus modified respect to the 
ground state only for the internal excitation energy ($\epsilon$) 
coming from the occupation of continuum states, which are treated with a 
Fermi gas level density parameterization. 
To avoid double counting of the free particles states \cite{bonche},
a high energy cut-off ($\tau$=9 MeV) is applied to the level density. 

To allow a comparison with the well known nuclear matter thermodynamics 
\cite{muellerserot,leemekjian,gulminelli}
and best isolate isospin effects on the fragmentation process,
we ignore the long range Coulomb interaction and,
in order to avoid interference with finite size effects,
consider equal size systems which differ by the neutron/proton ratio.

\section{Phase diagram}

The isospin dependence of the phase diagram for the MMM model is easily 
spotted in the microcanonical "isobar" ensemble, where energy is fixed and 
volume fluctuations are allowed and controlled through a conjugated Legendre 
intensive. 

Indeed, contrary to ordinary (macroscopic) thermodynamics, 
the thermal and phase characteristics
of a finite system depend on the statistical ensemble considered.  
The liquid-gas phase transition has a non-zero latent heat and has 
density as an order parameter,
meaning that the two associated phases can be distinguished by their different 
particle and energy densities. 
If a finite system (with a given fixed particle number) 
exhibits this transition, 
its event distribution at the transition point will show the two peaks 
corresponding to the two phases if and only if 
both energy and volume are free to fluctuate,
i.e. in the canonical isobar ensemble \cite{cg-phd}.

The general relationship between a distribution in the ensemble 
characterized by an intensive variable $\gamma$
associated to the conjugated extensive variable $m$,
and the Boltzmann entropy $W(m)=\exp S(m)$,
\begin{equation}
P_\gamma(m)= Z_\gamma^{-1} \exp \left ( S(m) -\gamma m \right ),
\end{equation}
insures then that at the liquid-gas transition point the 
compressibility is negative 
in the canonical isochore $(\beta, V)$ ensemble, and the heat capacity is negative in the 
microcanonical isobar $(E, P)$ ensemble. 

In the multifragmentation transition described by MMM, 
the low multiplicity ordered phase (compound nucleus)  
and high multiplicity disordered phase (multifragmentation)  
can be distinguished by their energy and volume,
just like in regular liquid-gas, and all the above considerations apply
\cite{npa2002}.
  
\begin{figure}
\resizebox{0.6\textwidth}{!}{%
  \includegraphics{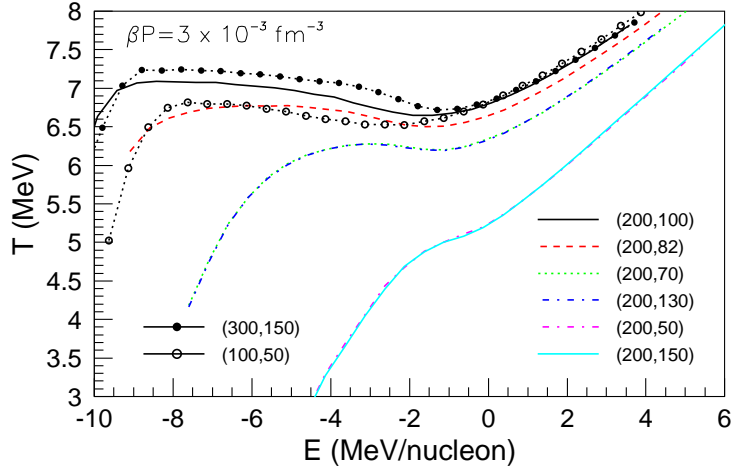}
  }  
  \caption{(Color online)
Microcanonical caloric curves at constant 
$\lambda=(\beta P)=3\cdot 10^{-3} {\rm fm}^{-3}$ for 200-nucleon systems 
and various neutron-proton ratios as indicated in the legend.
  $E$ is the total energy of the system.
  The magnitude of finite size effects with respect to isospin ones may be 
estimated considering the caloric curves with the same value of $\lambda$
 corresponding to the symmetric nuclei (100,50) and (300,150).
  }
  \label{fig:cc_constbP}
\end{figure}

Fig. \ref{fig:cc_constbP} shows
some constant $\lambda$ microcanonical 
caloric curves of a 200-nucleon systems with different
neutron-proton ratios.
The Lagrange parameter $\lambda$ can be associated
to a pressure through $P=\lambda T$, where 
$T=\left(\partial \bar{S}_E\left [\lambda \right ] / \partial E \right)^{-1} $ 
is the constant $\lambda$ microcanonical temperature.
The expected isospin invariance in the absence of Coulomb 
is confirmed by the fact that mirror nuclei 
((200,70) vs. (200,130),  and (200,50) vs. (200,150))
show an identical thermodynamical behavior, translated into
fully superimposable caloric curves.
The symmetric system shows a broad back-bending, 
signaling a liquid-gas-like phase transition.
Increasing the isospin asymmetry, the temperature shows a monotonic decrease,
and the back-bending width shrinks. 
The $\lambda$ value $\lambda = 3 \cdot 10^{-3}$ fm$^{-3}$, 
which still corresponds to the coexistence region for the $Z/A=0.35$ system,
appears clearly super-critical for $Z/A=0.25$.

The presence of Coulomb effects in physical nuclear systems 
breaks the n-p invariance and can considerably mask the 
isospin effects shown by Fig. \ref{fig:cc_constbP} \cite{npa2002}. 
In order to disentangle these effects, most of the experimental 
analyses concentrate on different isotopes of the same element 
(eg. $^{112}$Sn+$^{112}$Sn and $^{124}$Sn+$^{124}$Sn collisions).
In this case however, the sources have different total sizes and 
a naturally rising question is to what extend finite size effects interfering
with asymmetry effects may blur the signals of the last ones.
A quantitative answer is offered by Fig. \ref{fig:cc_constbP},
where caloric curves corresponding to
symmetric systems 50\% larger and 50\% smaller
than the previously discussed ones are considered.
The relative displacement of the curves suggests that for most of the 
presently analyzed
multifragmentation reactions finite size effects are small enough to 
be safely negligible.

\begin{figure}
\resizebox{0.5\textwidth}{!}{%
  \includegraphics{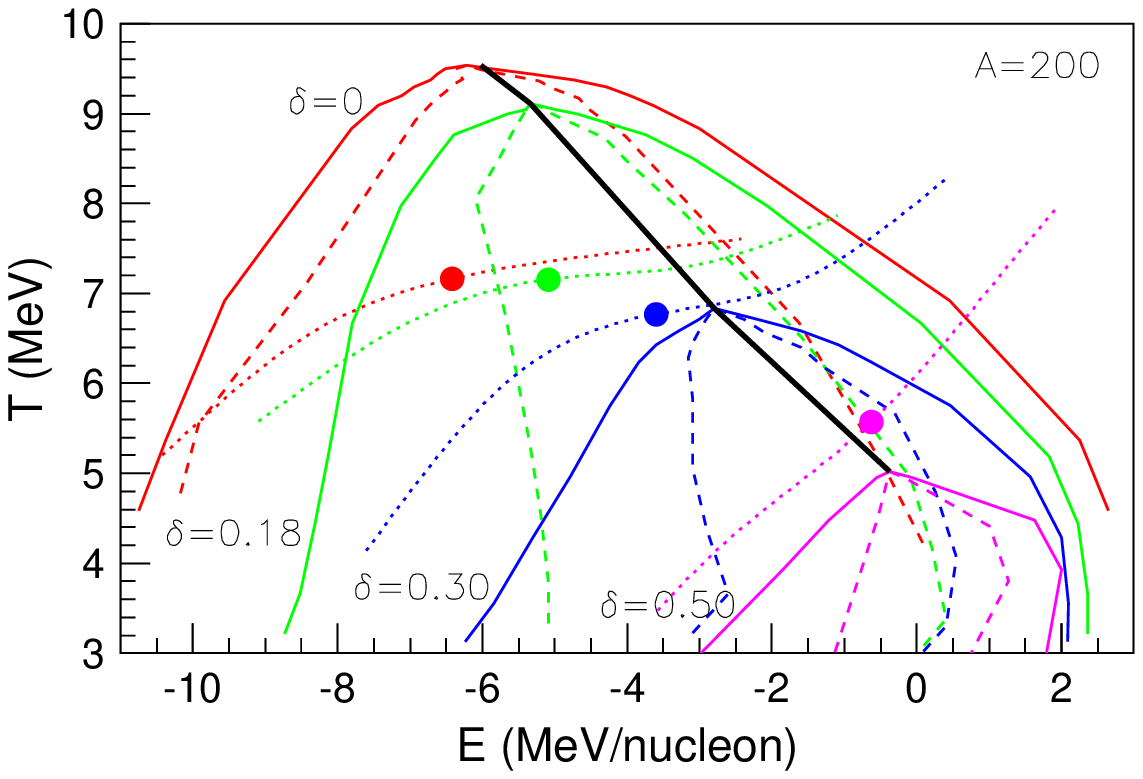}
  }
\resizebox{0.5\textwidth}{!}{%
  \includegraphics{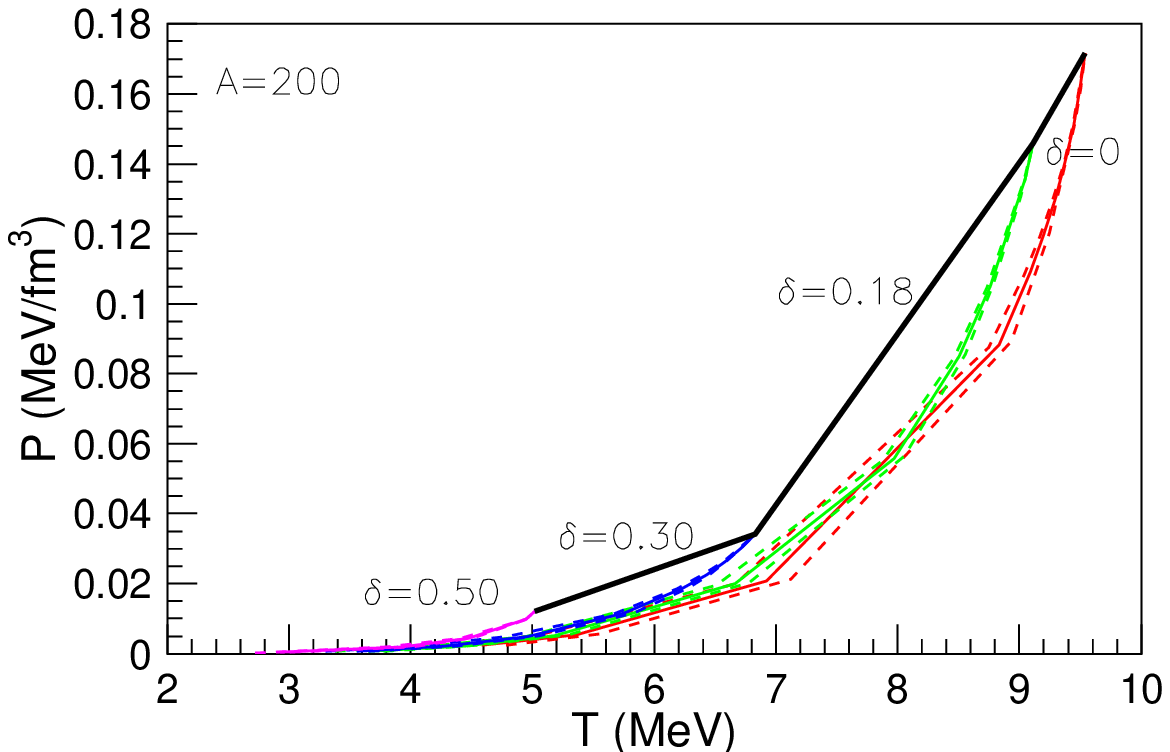}
  }
\resizebox{0.5\textwidth}{!}{%
  \includegraphics{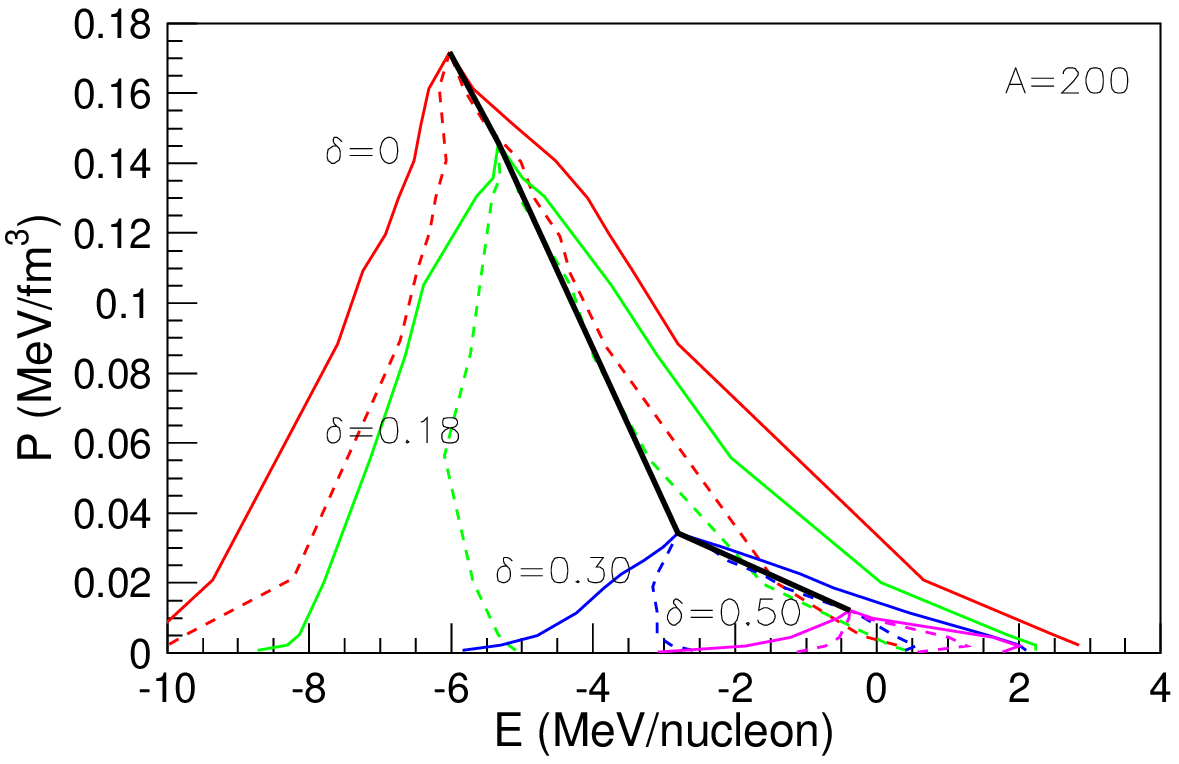}
  }
  \caption{(Color online)
    Projections in the temperature-total energy,
  pressure-temperature and pressure-total energy planes of the
  phase diagrams of 200-nucleon systems with various isospin asymmetries 
  ($\delta=(N-Z)/A$=0, 0.18, 0.30 and 0.50).
  The solid lines mark the borders of the coexistence region 
  while the dashed lines indicate the borders of the spinodal zone.
  The dotted lines in the upper panel show
  the paths followed through the phase diagram by the considered systems
  when the average freeze-out volume is fixed to 6$V_0$ and
  the excitation energy ranges from 2 to 10 MeV/nucleon.
  The full symbols point the location of multifragmentation events with 
  $E_{ex}$=6 MeV/nucleon further discussed in the text.
   }
\label{fig:phd} 
\end{figure}

The monotonic decrease of the critical temperature and 
pressure 
with the isospin asymmetry,
together with the reduction of the coexistence region, are illustrated in
Fig. \ref{fig:phd}, where the phase diagrams of the
200-nucleon systems with different asymmetries
are projected in the temperature-total energy,
pressure-temperature and pressure-total energy planes.
The solid lines correspond to the borders of the coexistence region, 
obtained from a Maxwell construction on the constant
$\lambda$ microcanonical caloric curves.
The dashed lines indicate the borders of the spinodal zone, defined 
by the back-bending extension for each $\lambda$ value.
The thick solid line connects the critical points in any representation.
By extrapolating this line, we can see that pure neutron and proton systems
may exist in the super-critical phase only, as expected from their inability to
form clusters at any temperature. It is interesting to remark 
that this intuitive result is obtained only as a limiting situation, 
while a phase transition survives with a sizeable critical temperature 
for systems as asymmetric as $Z/A=0.25$.

These results are in qualitative agreement with nuclear matter calculations
\cite{muellerserot,leemekjian,gulminelli},
showing that the thermodynamics of fragmentation of a finite nuclear system 
can be associated to the phenomenology of the nuclear matter 
liquid-gas phase transition.
This intuitive connection is systematically pushed forward in 
experimental studies,
however from a theoretical point of view this is not a trivial issue. 
If caloric curves and heat capacities in the
Statistical Multifragmentation Model (SMM) are available since more than two decades, 
most calculations with finite systems \cite{smm} were performed 
at constant volume and not constant pressure, and
therefore lead to signals that cannot univocally be interpreted as a 
first order phase transition.
A detailed exploration of the fragmentation phase diagram was presented 
in Refs. \cite{Das-PhysRep,bugaev}.
The results show a continuous transition from a high temperature
single phase to a mixed phase, this latter extending over the whole 
density domain of validity of the model \cite{bugaev}. 
This is reminiscent of the liquid-gas phenomenology,
but the transition from the liquid side, which is the only one accessible
experimentally, cannot be studied within this analytical model, and is 
hereby presented for the first time. The isospin dependence of the 
fragmentation phase diagram has never been studied before to our knowledge either.

\section{Isotopic distributions and fractionation}

In this section we turn to the connection between the system
phase diagram and cluster observables.

The generic feature of a first order fluid phase transition with two conserved 
particle numbers is the fractionation phenomenon: if the coupling between 
alike particles is less attractive then the coupling between not alike ones,
the ordered phase is systematically more symmetric than the disordered one. 
In the previous section we have shown that the fragmentation transition 
is qualitatively similar to such a fluid transition, we can therefore expect
to find a trace of fractionation in the fragments and particles chemical 
composition. 

Looking at the isotopic composition of fragments of 
different size emitted by neutron rich nuclei, it has been observed
that the isospin ratio $A/Z$ is a monotonically decreasing function 
of the fragment size, as one would intuitively expect if 
fractionation takes place and thermodynamic discontinuities 
are rounded by finite size effects. 
This "fractionation" phenomenon has been observed 
in the experimental analyses \cite{xu,geraci,martin,shetty,botvina},
but also in different dynamical models \cite{baoanli,ono,smf},
where it does not seem always connected to the phase coexistence
phenomenology. 

As we have stressed in section \ref{intro},
when dealing with finite systems the properties of coexisting phases 
cannot be deduced from the properties 
of pure phases by a simple linear combination . 
It is therefore not clear whether the neutron (proton)
enrichment of the nuclear gas (liquid) characteristic of phase 
coexistence in neutron rich nuclear 
matter \cite{muellerserot,leemekjian,gulminelli}
will be apparent in the partitions of the finite system inside
the coexistence region.

\begin{figure}
\resizebox{0.45\textwidth}{!}{%
  \includegraphics{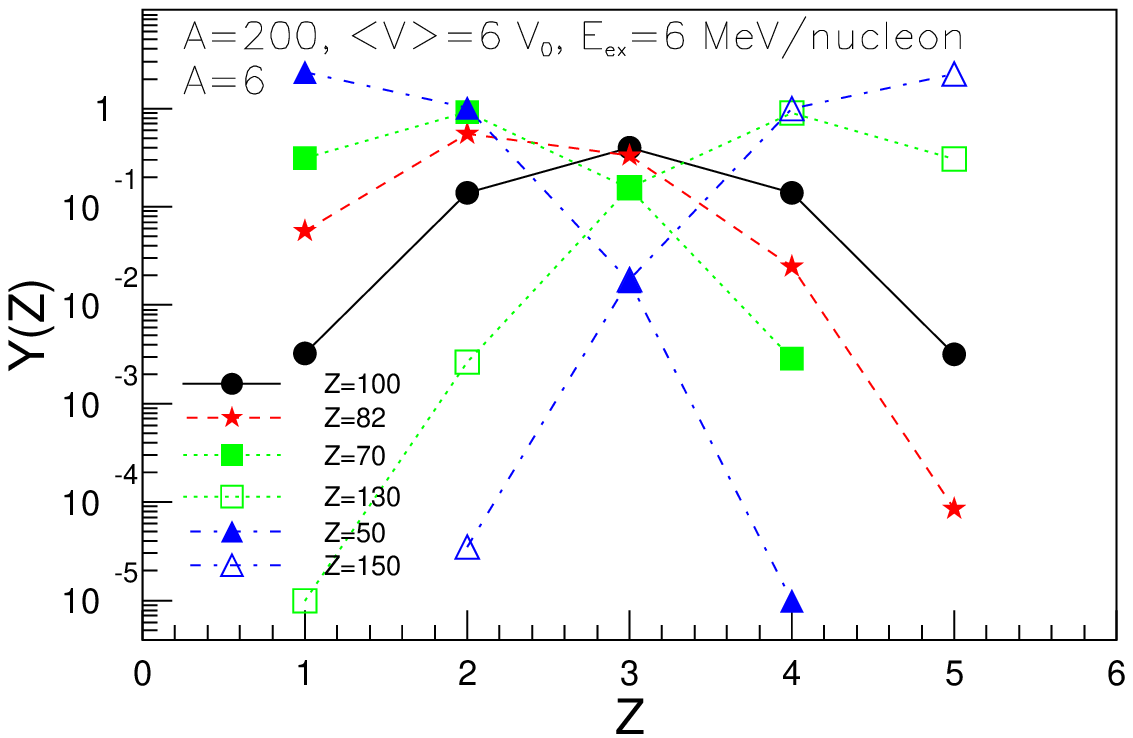}
  }
\resizebox{0.45\textwidth}{!}{%
  \includegraphics{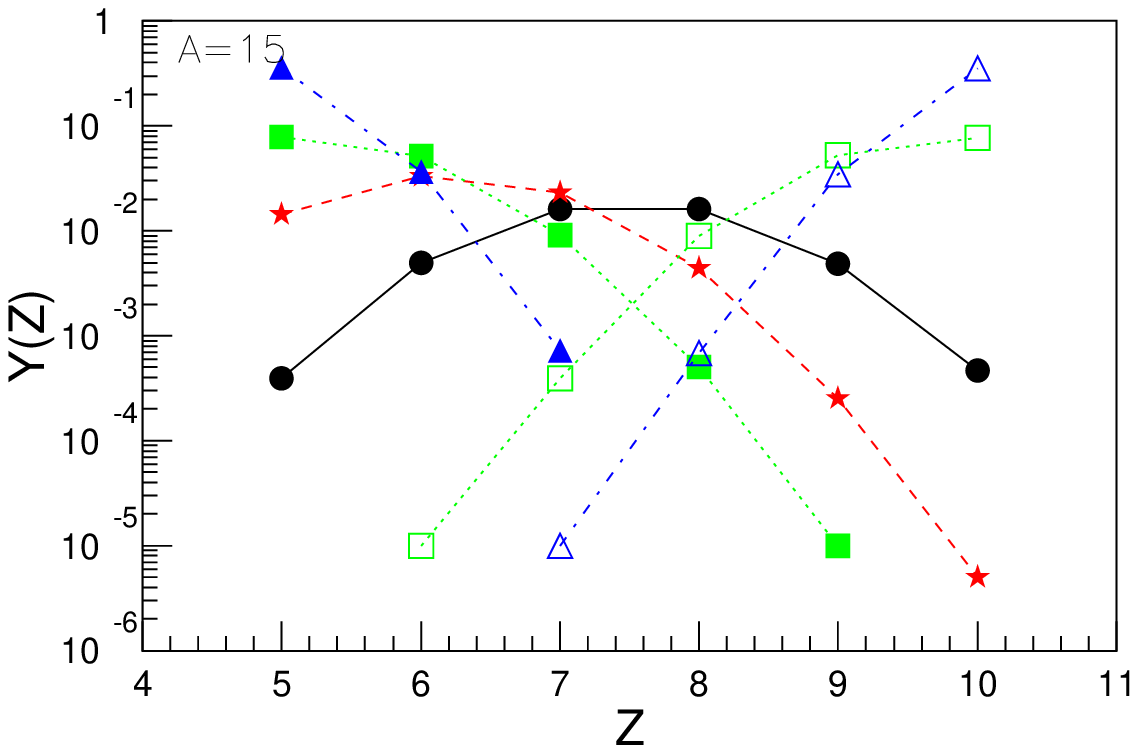}
  }
\resizebox{0.45\textwidth}{!}{%
  \includegraphics{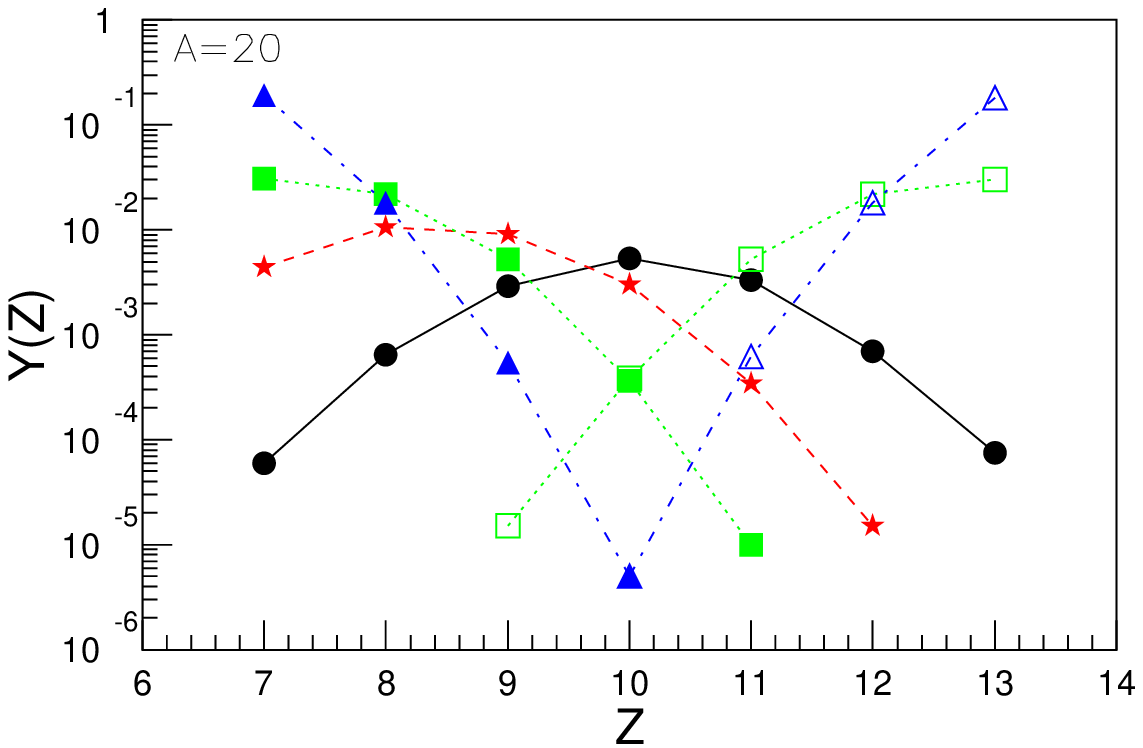}
  }
\resizebox{0.45\textwidth}{!}{%
  \includegraphics{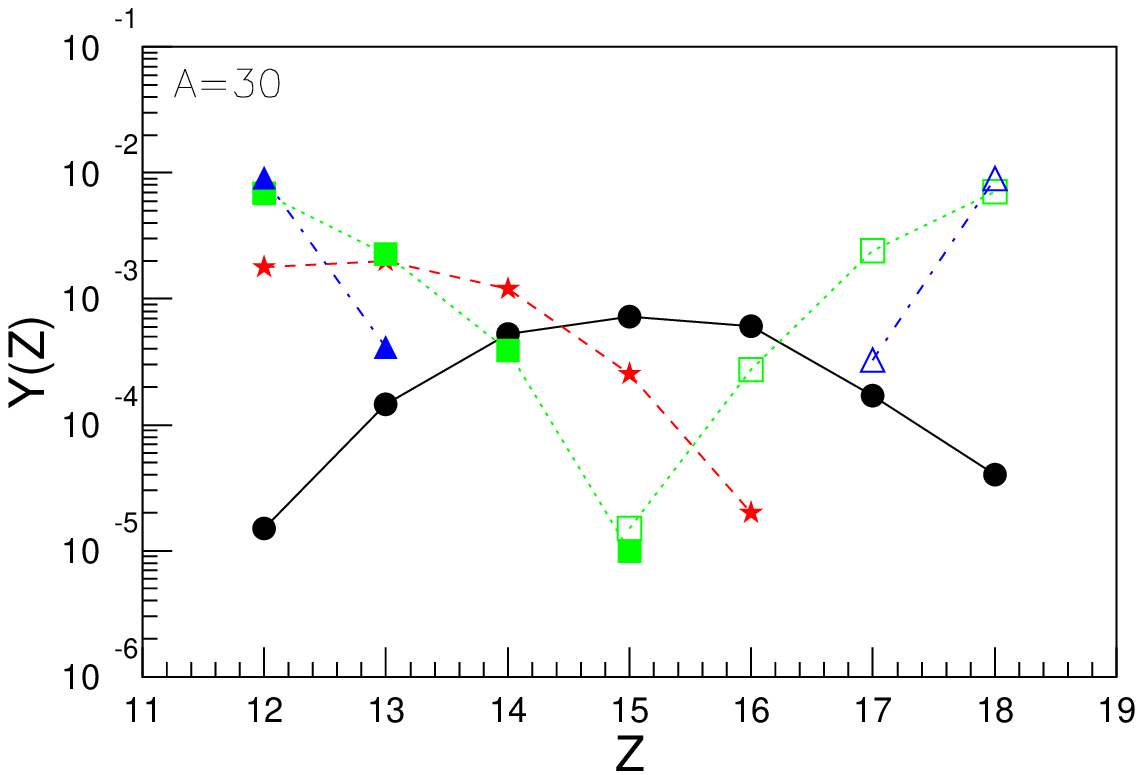}
  }
\caption{(Color online)
Break-up isotopic yield distributions of isobars with
$A$=6, 15, 20 and 30 originating from the multifragmentation of
200-nucleon sources with different isospin asymmetries
at 6 MeV/nucleon excitation energy and an average freeze-out volume
$\langle V\rangle =6 V_0$.}
\label{fig:y_iso}
\end{figure}

Fig. \ref{fig:y_iso} presents isotopic yield distributions of isobars with
$ A= 6, 15, 20, 30 $ obtained in the multifragmentation of nuclear systems
(200,100) ($\delta=(N-Z)/A$=0), (200,82) ($\delta$=0.18),
(200,70) ($\delta$=0.30), (200,50) ($\delta$=0.50), (200,130)
($\delta$=0.30) and (200,150) ($\delta$=0.50)
in a state representative for most multifragmentation reactions,
$E_{ex}$=6 MeV/nucleon and $\langle V\rangle =6 V_0$, 
where $V_0$ is the volume corresponding to normal nuclear density,
while the free neutron and proton multiplicities are listed in Table 
\ref{table:np_yield}.
These states are located in very different regions of the phase diagram. 
This is shown in the upper panel of Fig. \ref{fig:phd}, where
the dotted lines mark the paths followed
by the considered systems when the average volume is fixed to $\langle V\rangle =6 V_0$
and the excitation energy increases from 2 to 10 MeV/nucleon.
We can see that the $\langle V\rangle =6 V_0$ and $E_{ex}$=6 MeV/nucleon state 
(reported by full circles) is situated
well inside the spinodal zone for $\delta$=0 and $\delta$=0.18,
while it is close to the critical point for $\delta$=0.30 and
belongs to the super-critical region for $\delta$=0.50.

\begin{table}
\begin{center}
\begin{tabular}{ccccccc}
\hline
\hline
Multiplicity/Source & (200,100) & (200,82) & (200,70) & (200,50) & (200,130) & (200,150) \\ 
\hline
neutron & 6.71 & 1.31 $\cdot 10^1$ & 2.09 $\cdot 10^1$ & 3.27 $\cdot 10^1$ & 1.28 & 1.80 $\cdot 10^{-1}$\\
proton & 6.71 & 3.10 & 1.27 & 1.80 $\cdot 10^{-1}$ & 2.07 $\cdot 10^1$ & 3.24 $\cdot 10^1$ \\
\hline
\hline
\end{tabular}
\caption{Neutron and proton break-up multiplicity for different 200-nucleon
systems with an excitation energy of 6 MeV/nucleon and
an average freeze-out volume $\langle V\rangle =6 V_0$.}
\end{center}
\label{table:np_yield}
\end{table}

The distributions of Fig. \ref{fig:y_iso} exhibit some trivial
characteristic features:
the isospin symmetric source produces preferentially
isospin symmetric break-up fragments
(the isotopic yield distributions have a maximum at $Z=A/2$), 
and break-up fragment formation is invariant to n-p inversion 
(neutron and proton yields are equal and isotopic yield distributions are
symmetric with respect to $A/2$).
Concerning the isospin asymmetric systems, one can see that the
more neutron (proton) rich is the source, 
more free neutrons (protons) are emitted and
neutron (proton) richer are the break-up fragments.
The isospin invariance in the absence of Coulomb is confirmed by the
isotopic yield distribution of the mirror nuclei (200,70) vs. (200,130)
and (200,50) vs. (200,150), which have reflection symmetry with respect
to the $Z=A/2$ axis.

\begin{figure}
\resizebox{0.6\textwidth}{!}{%
  \includegraphics{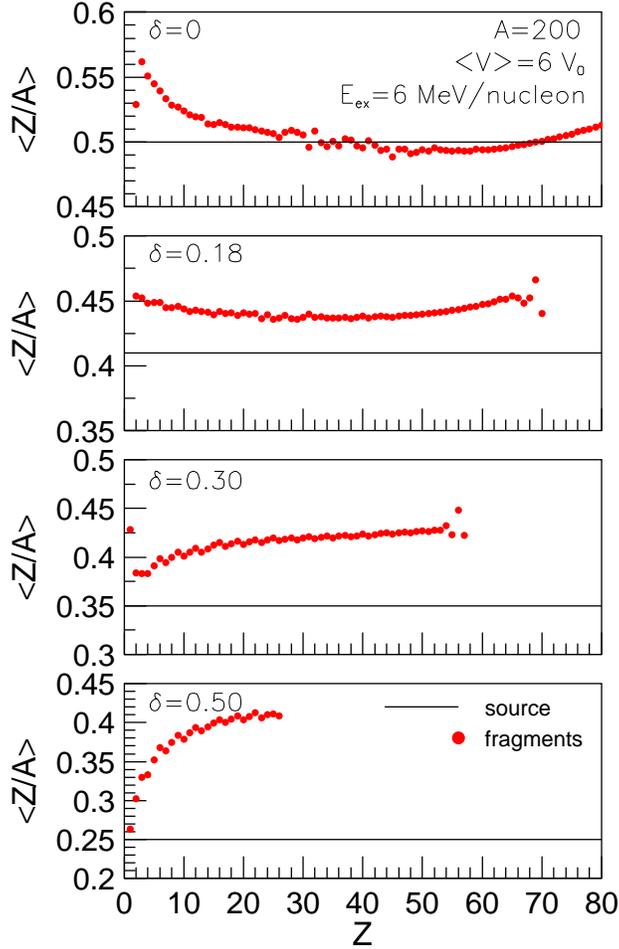}
  }
  \caption{(Color online)
    Average isospin content ($Z/A$) of the break-up fragments
  produced in the multifragmentation of different neutron rich
  200-nucleon systems with asymmetries
  $\delta$=0, 0.18, 0.30, 0.50 as a function of fragment charge.  
  In all cases $\langle V\rangle =6V_0$ and $E_{ex}$=6 MeV/nucleon.
  The horizontal solid lines indicate the isospin content of the sources.
  }
\label{fig:zova}
\end{figure}

One can also notice that the distributions are cut both on the proton rich
and on the neutron rich side. This is due to the dramatic decrease
of the binding energy with isospin asymmetry approaching the drip-lines.
As a consequence, for asymmetric sources
primary fragments tend to be more symmetric than the initial source,
the total asymmetry of the system being preserved by a correspondingly
increased number of free neutrons (protons).
The fractionation induced by this effect can be appreciated from 
Fig. \ref{fig:zova}, which shows as a function of the fragment size its average
isospin content for the four different asymmetries considered above.

The first feature arising from Fig. \ref{fig:zova} is that fragments are 
usually more proton-rich than the corresponding source. 
The only exception corresponds to the symmetric source and
fragments whose charge is close to half the source charge, 
where mass and charge conservation induces for the $Z/A$ ratios
values slightly lower than 0.5. 
Thus, the approximation frequently invoked in multifragmentation studies, that 
primary fragments have the same $N/Z$ of their 
emitting source \cite{tsang,botvina,schmidt,palluto}, does not seem to 
be correct, if primary partitions correspond to statistical equilibrium.

Even more important, the degree of fractionation is seen to monotonically increase 
with the asymmetry of the source, independent of the location 
of the multifragmentation event in the phase diagram.
Indeed, the occurence of fractionation directly follows from the
isospin content of the free particles 
because of mass and charge conservation and, thus, it 
cannot be taken as a signature of coexistence in finite systems.
On the other hand, the behavior of fractionation with the fragment size 
is very different depending on the thermodynamic characterization of the system.
Inside the spinodal region (two upper panels),
the $\langle Z/A\rangle $ vs. $Z$ distributions 
shows a clear "U" shape 
which has been already discussed in Ref. \cite{isospinfrac}.
The two bottom panels which refer to system in a "pure" phase (liquid or fluid) 
present similar characteristics which are very different from the behavior discussed
above: both show a monotonically increasing $\langle Z/A\rangle $ vs. $Z$ distribution
which for $Z_{source}/4$ reaches saturation at about $\langle Z/A\rangle \approx$0.42. 

These observations mean 
that the fractionation phenomenon naturally appears
as soon as the fragmentation process is ruled by thermal laws.
It allows to identify the coexistence region
of the first order phase transition only if an accurate isotopic
characterization of all emitted fragments is possible.

\begin{figure}
\resizebox{0.45\textwidth}{!}{%
  \includegraphics{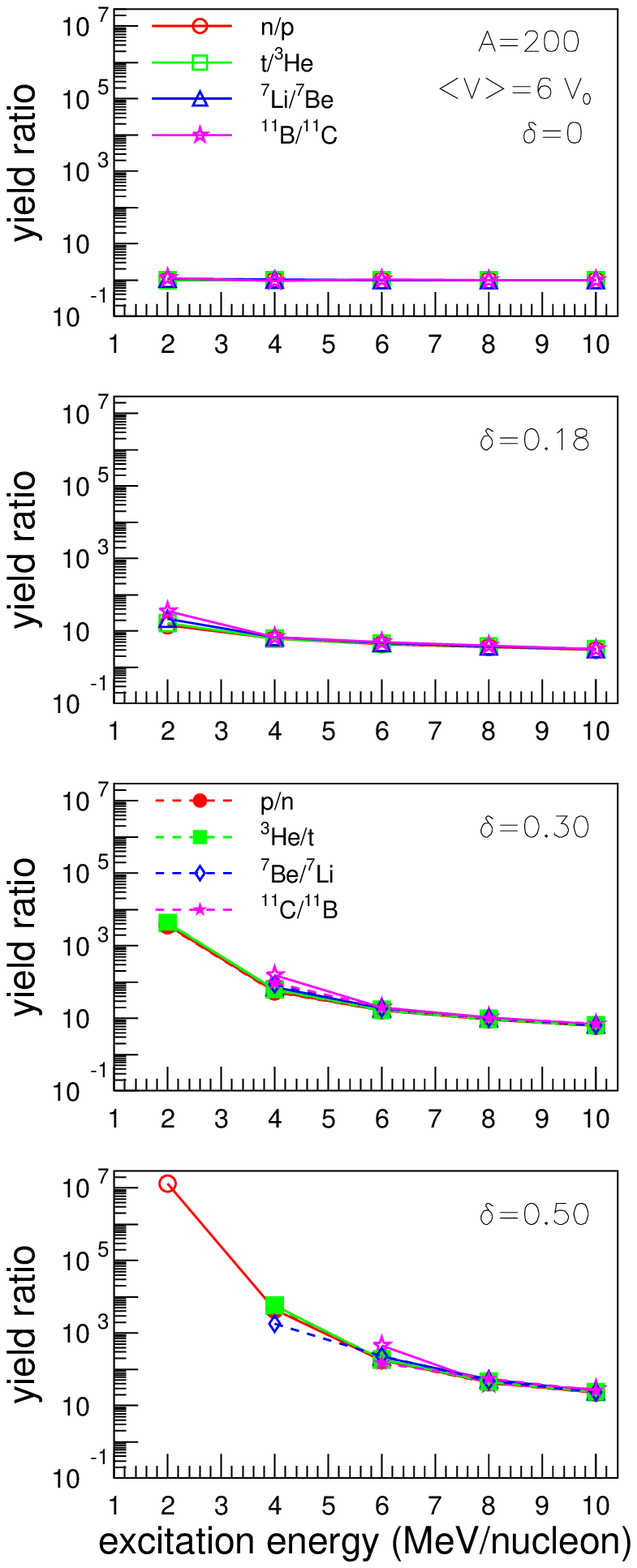}
  }
  \caption{(Color online)
    Ratios of isotopic yields of different light mirror nuclei
  at the break-up stage of 200-nucleon systems with different isospin
  compositions as a function of excitation energy.
  In all cases the average freeze-out volume is $\langle V\rangle =6 V_0$. 
  For the neutron-rich sources the considered ratios are with the
  neutron-rich isobars in the numerator while for the neutron-deficient
  sources the neutron-rich isobars are in the denominator.}
\label{fig:rap_ex}
\end{figure}

The energy and asymmetry dependence of fractionation 
is further explored in Fig. \ref{fig:rap_ex},
which gives the evolution with excitation energy of
various ratios of light mirror nuclei isotopic yields.

The top panel corresponds to the symmetric source (200,100) and the results
indicate that, no matter the excitation energy, mirror nuclei are produced
with equal probability.
The lower panels show that this is not true in asymmetric systems,
and the increase in the emission probability for asymmetric light clusters
respect to combinatorial expectations, increases with the asymmetry of the
source.
Similar to the results of Fig. \ref{fig:zova}, we can see that fractionation is
mainly dictated by the number of evaporated nucleon in excess,
and no special pattern may be distinguished for the events located inside
the coexistence region respect to those situated in the liquid or supercritical
regimes. Indeed, similar values are obtained, for instance, for
the $\delta=0.18$ source with $E_{ex}$=4 MeV/nucleon (phase coexistence region)
and the $\delta=0.30$ source at $E_{ex}$=8 MeV/nucleon (supercritical region).
Concerning the dependence with excitation energy, we can see that 
the energy increase partially washes out  the trend of the neutron rich systems
to preferentially produce neutron rich fragments.
This is in agreement with statistical microscopic models \cite{FGisospin}.

Based on the idea of isospin fractionation, it has been proposed that the gas
neutron enrichment can be measured from such ratios \cite{tsang,geraci}.
Indeed, in the grand-canonical approximation, if the charge difference between
the two isobars is $\Delta Z=1$, the isobaric ratio is given by 
\begin{equation}
R=\frac{Y(A,Z_1)}{Y(A,Z_2)}= \exp \left(\frac{\Delta B +\Delta \mu}{T} \right),
\end{equation}
where $\Delta B=B(A,Z_1)-B(A,Z_2)$, and $\Delta \mu=\mu_n-\mu_p$, 
if the neutron rich isobar is in the numerator. 
Choosing light isobars having close binding energies, this ratio is then a
direct measure of the chemical potential difference between neutrons and
protons, i.e. of the ratio $\rho_n/\rho_p$ of the free neutron-proton
densities.   

Fig. \ref{fig:rap_ex} shows that,
 at least for the considered fragments,
this approximation is reasonable enough for
all excitation energies and asymmetries,
and information on the free neutron versus proton behavior can
indeed be inferred from the measurements of isotopically 
resolved light fragments, assuming that freeze-out yields 
can be restored from the experimentally detected cold fragments.
However, 
it is important to stress that this free densities ratio cannot 
unambiguously sign the neutron enrichment of the gas phase, 
since, as discussed above, the same behavior is observed 
when no gas phase can be thermodynamically defined.

To conclude, in this section we have pointed out that isospin
fractionation cannot be taken as a signature of phase coexistence
when dealing with finite systems.
In a previous work \cite{npa2002}, we have already shown that the Coulomb
interaction tends to quench the coexistence zone: because of that
the multi-fragmentation phenomenology can be associated 
with a super-critical region of the charged-system phase diagram.
Here we show that, even in the absence of the Coulomb interaction,
a generic universal feature of fragmentation, namely isospin
fractionation, can show-up above the critical point.
If the effect of Coulomb on the phase diagram strongly depends
on the specific model used \cite{npa2002,ison}, this super-critical fractionation
is a generic effect, which we believe should be present in any fragmentation
model. Indeed it is due to the combined effect of clustering 
in the super-critical region (which favors the formation of 
fragments close to stability, $i.e.$ an isospin symmetric fraction of the 
system condensed at finite baryon density),
and particle number conservation (which forces the low density non-clustered
part to have a strong isospin asymmetry). 
These features are naturally present in any model, microscopic or macroscopic, 
respecting conservation laws and ruled by the competition
between entropy and energy.
Such an effect is not accessible in nuclear matter calculations, where
the sharing of the system into a dense and a diluted fraction is by construction
a sign of phase separation. 
Classical models \cite{campi} have already shown clustering in the supercritical region,
but this is to our knowledge the first time that 
this effect is reported in a realistic nuclear multifragmentation model.

\section{Isotopic widths and symmetry energy  in finite neutral systems}

A very powerful motivation in the study of isotopic distributions in
fragmentation  reactions is given by the well-spread expectation that
information coming from the low
density finite temperature coexistence zone of the phase diagram will be 
sensitive to the symmetry energy coefficient of the nuclear (free) energy -
density functional at finite temperatures, and densities well below saturation
\cite{tsang,LeFevre,shetty04,Tsang2,Ono2,baoan06}. 
Such analyses would then be complementary to isospin diffusion and 
neutron skin measurements \cite{steiner}
and additionally give a unique information on temperature effects on the 
symmetry energy.

In the MMM model the binding energy of a cluster
of mass $A$ and charge $Z$ is parameterized as,
\begin{eqnarray}
B_{noC}(A,Z)&=&(a_v A - a_s A^{2/3}) - a_i(a_v A - a_s A^{2/3}) \frac{(A-2Z)^2}{A^2}
\nonumber \\
&=& (a_v A - a_s A^{2/3}) - C_{sym}(A) \frac{(A-2Z)^2}{A},
\label{eq:bld}
\end{eqnarray}
and includes a full mass dependence of the bulk+surface and
isospin-dependent contributions.
The Coulomb part of the binding energy ($a_c Z^2/A^2+a_a Z^2/A$)\cite{ld} 
which is included in the standard version of MMM \cite{mmm},
is switched off for this study, to concentrate on isospin effects.

As we have stressed in section \ref{theory},
in the framework of statistical models under the Fisher approximation 
\cite{fisher},
the low density correlations are entirely exhausted by clusterization.
This means that the symmetry (free) energy entering in the fragment production
yields inside coexistence should be the symmetry energy of isolated nuclei at finite temperatures. 
In particular the interaction part of this energy (Eq. (\ref{eq:bld})) should 
correspond to normal ground state values, meaning that 
the liquid drop parameters $a_i,a_v,a_s$ have standard ground state values \cite{ld}. 
In our model isotopic yields are therefore expected to be entirely determined
in the whole phase diagram by the symmetry energy coefficients $C_{sym}(A)$.

If this approximation gives a correct description of multifragmentation,
this would mean that no relevant information on $C_{sym}(\rho,T)$ can be inferred 
from fragment observables.
If, on the other hand, the energy functional of break-up fragments
differs from the ground-state functional \cite{LeFevre,shetty04},
it would be interesting to trace its behavior and isotopic distributions
could be good candidates. The extent to which this may be true is a difficult 
theoretical issue, demanding a quantal many-body transport treatment, 
completely out of scope of the present study. 
Whatever the final answer, for the fragment symmetry-energy to be 
accessible from experimental data, it is necessary to prove that 
in a controlled model where the symmetry energy set in break-up fragments
is an input value of the calculation, the proposed observables do indeed recover 
its value within a good precision.

The information on $C_{sym}$ can be directly inferred from the widths of the 
isotopic distributions in the grandcanonical approximation.
Indeed, a Gaussian approximation on the grandcanonical expression,
\begin{equation}
Y_{\beta,\mu_n,\mu_p}(N,Z)
={\cal Z}_{\beta,\mu_n,\mu_p}^{-1} 
\exp \left[ -\beta \left( F_{\beta}(N,Z) - \mu_n N -\mu_p Z\right ) \right],
\label{eq:grancano}
\end{equation}
leads to
\begin{equation}
Y_{\beta,\mu_n,\mu_p}(A,N-Z) = K(A) \exp \left[ -  
\frac{ \left( N-Z -I_0 \right )^2}{2 \sigma_I^2(A)} \right],
\end{equation}
where $I_0=\overline{N}-\overline{Z}$ is the most probable value of $N-Z$ 
for a given value of the cluster size $A$, 
$K(A)$ does not depend on the asymmetry $I=N-Z$ and the isospin variance is
related to the symmetry energy coefficient by,
\begin{equation}
\sigma_I^2(A)= \frac{AT}{2C^{\beta}_{sym}(A)}.
\label{eq:csym_fluct}
\end{equation}

\begin{figure}
\resizebox{0.6\textwidth}{!}{%
  \includegraphics{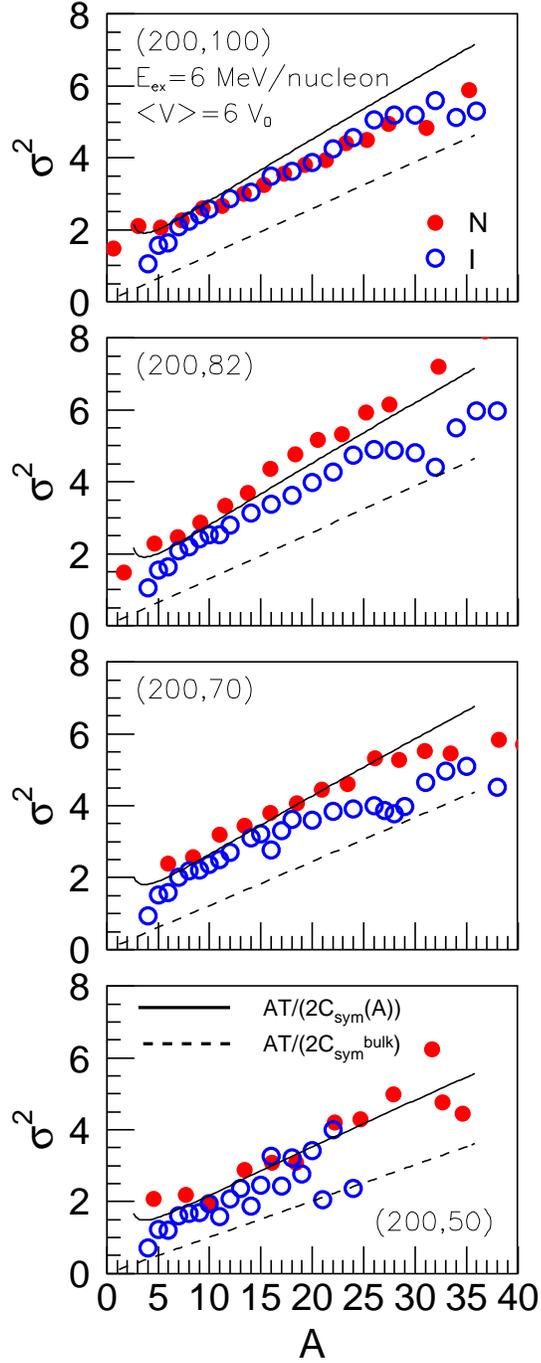}
  }
  \caption{(Color online)
    Widths of $Y(I)|_A$ (open symbols) and $Y(N)|_Z$ (full symbols)
  distributions corresponding to the break-up stage of 200-nucleon systems
  with different asymmetries ($\delta$=0, 0.18, 0.30, 0.50) as a function of
  fragment mass in comparison with predictions of Eq. (\ref{eq:csym_fluct_fr})
  calculated for fragment (full line) and bulk (dashed line)
  symmetry energies. 
  All systems are characterized by 
  $\langle V\rangle =6V_0$ and $E_{ex}$=6 MeV/nucleon
  and the Coulomb interaction is switched off.}
\label{fig:sig2}
\end{figure}

The coefficient $C^{\beta}_{sym}(A)$ appearing in this last expression is 
a free symmetry energy coefficient given by,
\begin{equation}
C^{\beta}_{sym}(A)=\frac{A}{2} 
\frac {\partial^2 F_{\beta}(N,Z)}{\partial I^2}|_A,
\end{equation}
and coincides with $C_{sym}(A)$ defined by Eq. (\ref{eq:bld})
if we neglect the $I$ dependence of the excitation energy and entropy
associated to a given mass $A$.
In this case Eq. \ref{eq:csym_fluct} reads,
\begin{equation}
\sigma_I^2(A) \approx \frac{AT}{2C_{sym}(A)}.
\label{eq:csym_fluct_fr}
\end{equation}
The quality of all these approximations can be appreciated from 
Fig. \ref{fig:sig2}, which displays the width of the asymmetry distribution
(open circles)
as a function of the fragment mass for the same thermodynamic conditions as 
Figs. \ref{fig:y_iso}, \ref{fig:zova}, \ref{fig:rap_ex}.
For all considered asymmetries, Eq. (\ref{eq:csym_fluct_fr}) (solid line)
appears well verified for small masses, meaning that this width can 
indeed be taken as a measure of the underlying symmetry energy.
For higher masses and/or extreme asymmetries the Gaussian approximation
breaks down as it can be observed from Fig. \ref{fig:y_iso}, 
and the link between symmetry energy and fluctuation is lost.
The neutron distribution (filled circles) contains approximately the same 
information brought by the asymmetry distribution.
The dashed lines in Fig. \ref{fig:sig2} give the grandcanonical expectation
for the isotopic widths Eq. (\ref{eq:csym_fluct_fr}), 
when only the bulk term ($C_{sym}^{bulk}=a_i a_v$)
of the symmetry energy is considered, instead of the complete expression
($C_{sym}(A)=a_i a_v-a_i a_s A^{-1/3}$) used for the solid line.
We can see that accounting for surface effects in the fragment energy functional
leads to a considerable increase of the isospin widths even for relatively
massive fragments. Such fragments ($A\geq 30$) are not adapted to study the
symmetry energy coefficient though, because of the important effect of the 
mass and charge conservation constraint that causes the widths to deviate
from their grandcanonical expectation.

It is very interesting to observe that the solid and dashed lines are
almost parallel to each other. This means that the expected functional dependence 
of the width on the mass number does not change drastically if 
the symmetry energy has a surface contribution or not.
As a consequence, a large width as expected for light, surface-dominated 
fragments ($A\leq 30$), can be easily mis-interpreted as a signature 
of a reduced bulk symmetry value, as it has suggested by some recent publications 
\cite{LeFevre,shetty04}.

The relation between isotopic widths and symmetry energy has interesting
consequences on the isoscaling observable, which has raised a great interest
in the recent literature \cite{ono,LeFevre,shetty04,baoan06}. 
As long as the distributions can be approximated
by Gaussians, the ratio between the production yield of the same isotope
in two different systems $(1)$ and $(2)$ 
(where we denote by (2) the neutron-rich one) for a given $Z$
can be expressed  as a function of neutron number $N$ as: 

\begin{equation}
\ln \left ( \frac{Y_{(2)}(N,Z)}{Y_{(1)}(N,Z)}\right ) = 
-\frac{N^2}{2} \left ( \frac{1}{\sigma^2_{N_{(2)}}}-
\frac{1}{\sigma^2_{N_{(1)}}} \right )
+ N \left ( \frac{\overline{N}_{(2)}}{\sigma^2_{N_{(2)}}}-
\frac{\overline{N}_{(1)}}{\sigma^2_{N_{(1)}}} \right )
+K(Z),
\label{eq:gaussian}
\end{equation}
where $\overline{N}_{(i)}$ is the most probable $N$ value for the element 
$Z$ in system $(i)$ and $\sigma^2_{N_{(i)}}$ is the variance of the
$N$ distribution in the same system. 
If we choose as systems $(1)$ and $(2)$ two systems of similar masses and
temperatures, then
$\sigma^2_{N_{(1)}}\approx\sigma^2_{N_{(2)}}\approx\sigma^2_{N}$
and the ratio shows (in log scale) a linear dependence on 
$N$ (at fixed $Z$).
Similar arguments hold also for fragments with fixed $N$, meaning that
the quantity in left-hand side of Eq. (\ref{eq:gaussian}) has also a
linear dependence on $Z$ (at fixed $N$). 
This result is known in the literature as the isoscaling phenomenon \cite{tsang}:
\begin{equation}
\ln \left ( \frac{Y_{(2)}(N,Z)}{Y_{(1)}(N,Z)}\right ) = \alpha(Z) N + K(Z), 
\label{eq:isoscaling}
\end{equation}
where,
\begin{equation}
\alpha(Z)=\frac1{\sigma_N^2} \left( \overline{N}_{(2)}-
\overline{N}_{(1)}\right),
\label{eq:alpha1}
\end{equation}
or, for symmetric distributions,
\begin{equation}
\alpha(Z)=\frac1{\sigma_N^2} \left( \langle N \rangle _{(2)}-\langle N\rangle _{(1)}\right).
\label{eq:alpha2}
\end{equation}

In the mass region where the Gaussian approximation is well verified and
in the absence of Coulomb effects (see Fig. \ref{fig:sig2}),
\begin{equation}
\sigma^2_N(Z)\approx\sigma^2_I(\langle A\rangle (Z)).
\label{eq:approx_sig2}
\end{equation}

Then, the isoscaling parameter $\alpha(Z)$ is linked to the fragment symmetry
energy by,
\begin{equation}
\alpha(Z) \approx \frac{2 C_{sym}(\langle A\rangle )}
{\langle A\rangle T}\left ( \langle N\rangle _{(2)}-\langle N\rangle _{(1)} 
\right ), 
\label{eq:isoscaling_fluct}
\end{equation} 
or, equivalently,
\begin{equation}
\alpha(<Z(A)>) \approx \frac{2 C_{sym}(A)}
{AT} \left ( \langle I\rangle _{(2)}-\langle I\rangle _{(1)} 
\right ). 
\label{eq:isoscaling_flucti}
\end{equation} 

\begin{figure}
\resizebox{0.8\textwidth}{!}{%
  \includegraphics{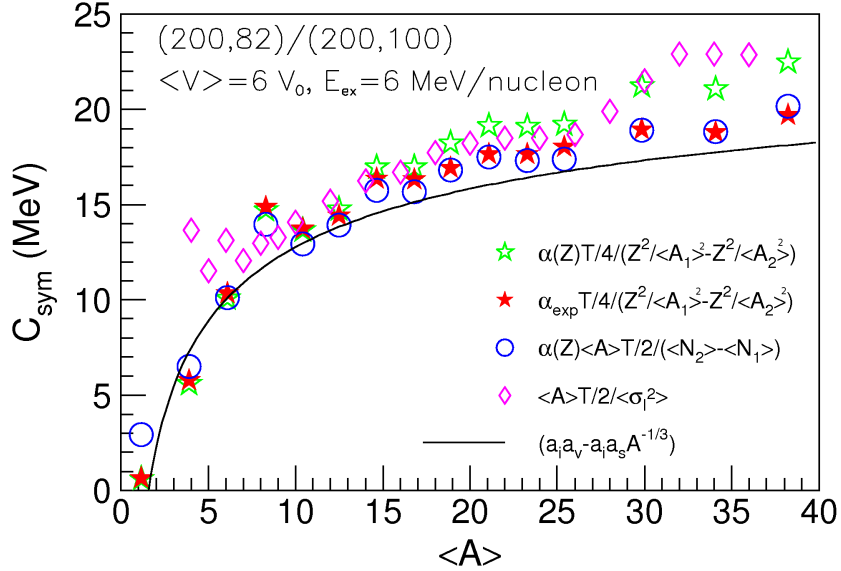}
  }
  \caption{(Color online)
    The input symmetry energy coefficient as a function of fragment size
  (solid line) is compared to the estimations from fragment observables,
  calculated for the break-up stage of (200,82) and (200,100) nuclear systems 
  without Coulomb with
  $\langle V\rangle =6V_0$ and $E_{ex}$=6 MeV/nucleon.
  Open diamonds and circles stand for predictions of 
  Eq. (\ref{eq:csym_fluct_fr}) and, respectively, Eq. (\ref{eq:isoscaling_fluct}). 
  Predictions of Eq. (\ref{eq:csym_ono}) (stars) are plotted as
  open or full stars depending whether 
  the isoscaling parameter $\alpha$
  is calculated according to its definition (Eq. \ref{eq:isoscaling}) or
  as the average values of fragments with $1 \leq Z \leq 8$.
}
\label{fig:casym_noc}
\end{figure}

Fig. \ref{fig:casym_noc} compares the symmetry energy coefficient extracted from
Eq. (\ref{eq:isoscaling_fluct}) 
with the input symmetry energy of the model for a representative case.
Different average volumes, excitation energies, asymmetry ratios give 
similar results. 
We can see that once again the Gaussian approximation appears well 
verified for light fragments and, in that case,
isoscaling technics give a reasonably good measure of 
the fragment symmetry energy, the reason being that this parameter is directly
linked to isotopic fluctuations.

A similar expression,
\begin{equation}
\alpha(Z) \approx \frac{4 C_{sym}(\langle A\rangle )}{T}
\left ( \frac{Z^2}{\langle A\rangle _{(1)}^2} - 
\frac{Z^2}{\langle A\rangle _{(2)}^2} \right ),
\label{eq:csym_ono}
\end{equation} 
has been derived in Ref. \cite{ono} from Eq. (\ref{eq:grancano}) 
with a similar saddle point approximation as for Eq. (\ref{eq:isoscaling_fluct}),
considering only the most probable isotopes for each $Z$.
This equation is also plotted on Fig. \ref{fig:casym_noc} and gives comparable 
results to Eq. (\ref{eq:isoscaling_fluct}).

In the some experimental analyses the isoscaling parameter $\alpha(Z)$ is 
not extracted separately for each isotope, as defined in Eq. (\ref{eq:isoscaling}),
but as the average value over fragments with $1\leq Z\leq 8$.
For this last situation we adopt the notation $\alpha_{exp}$. 
Fig. \ref{fig:casym_noc} plots the behavior of Eq. (\ref{eq:csym_ono})
for these different definitions of $\alpha$.
It comes out that the procedure to calculate $\alpha$ using exclusively
the light nuclei 
does not perturb the extracted symmetry energy, but rather minimizes the 
deformations due to conservation laws, which would prevent a precise extraction
of the symmetry energy coefficient for large clusters.
Quantitatively speaking, by ignoring the monotonic increase with $Z$, 
the use of $\alpha_{exp}$ results in slightly lower values of the
symmetry energy with respect to the ones obtained employing $\alpha(Z)$.

It has been recently argued \cite{Ono2,baoan06} that a constant value of the 
isoscaling parameter $\alpha$ would imply a bulk character for the associated
symmetry energy.
It is particularly interesting to notice that in our model this is not the
case. Indeed, the size dependence of $C_{sym}(\langle A\rangle )/\langle A\rangle $
is compensated by the size dependence of
$(\langle N\rangle _{(2)}-\langle N\rangle _{(1)})$ giving a constant $\alpha$.

\section{Isotopic widths and symmetry energy in real nuclei}

The robustness of these signals to measure the symmetry energy 
when the Coulomb interaction is included is particularly important
as it gives the extent to which one may extract this basic quantity
from multifragmentation data, assuming that one may 
access the chemical composition
of the physically relevant break-up fragments.

In ground state nuclei 
the Coulomb interaction is known to shift the stability peak towards neutron rich
nuclei and to reduce dramatically the binding energy of the proton rich ones.
This last effect is responsible for a strong narrowing of the 
$B(A,Z)|_A$ and $B(A,Z)|_N$ distributions and gets more pronounced
with the mass increase.
Since the logarithm of fragments multiplicity approximately follows the evolution of
$B(A,Z)/T$ (Eq. (\ref{eq:grancano})),
we can expect to find the same effect of a width reduction on fragments yields.

\begin{figure}
\resizebox{0.6\textwidth}{!}{%
  \includegraphics{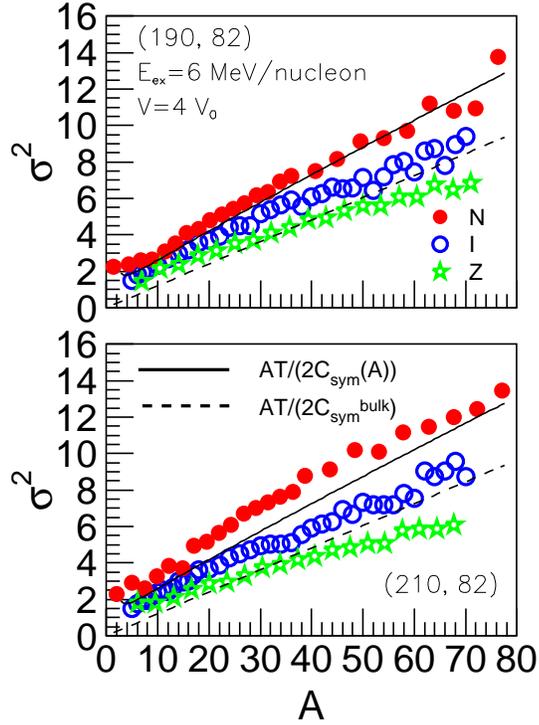}
  }
  \caption{(Color online)
    Widths of $Y(I)|_A$ (open circles), $Y(N)|_Z$ (full circles)
  and $Y(Z)|_N$ (open stars)
  distributions corresponding to the break-up stage of two $Z$=82 nuclei
  and different asymmetries, $\delta$=0.14 (upper panel) and 0.22 (lower panel)
  characterized by the freeze-out volume
  $V=4V_0$ and excitation energy $E_{ex}$=6 MeV/nucleon
  as a function of
  fragment mass in comparison with predictions of Eq. (\ref{eq:csym_fluct_fr})
  calculated for fragment (full line) and bulk (dashed line)
  symmetry energies.}
\label{fig:sig2+c}
\end{figure}

Fig. \ref{fig:sig2+c} plots the widths of the $Y(I)|_A$, $Y(N)|_Z$ and, 
for the sake of completeness, $Y(Z)|_N$ distributions for two nuclei,
(210, 82) and (190, 82), in a thermodynamical state relevant for
most multifragmentation reactions, $V=4 V_0$ and $E_{ex}$=6 MeV/nucleon.
The choice of a pair of  nuclei with equal charge minimizes the interference between
Coulomb and isospin contributions.

We can observe that the expected dispersion between $\sigma_I^2$,
$\sigma_N^2$ and $\sigma_Z^2$ increases with the fragment mass and the source asymmetry, 
which obviously restricts the validity of the approximation Eq. (\ref{eq:approx_sig2}).
The decrease of $\sigma_I^2$ and $\sigma_Z^2$ can be attributed to the
above mentioned narrowing of $B(A,Z)|_A$ and $B(A,Z)|_N$ distributions under Coulomb effect.
The increase of $\sigma_N^2$ then results from particle number
conservation. Indeed, under mass and charge conservation
the dispersion of fragment yields obeys the law:
\begin{equation}
\sigma_I^2 \approx \left(\sigma_N^2 + \sigma_Z^2\right)/2.
\end{equation}

The consequence of these effects is that the slight deviation of 
$\sigma_I^2$ from Eq. (\ref{eq:csym_fluct_fr}), already present for the heavy fragments
obtained in the decay of finite neutral systems,
gets more pronounced under Coulomb such that for 
$A>50$, $\sigma_I^2$ practically falls over the predictions of
Eq. (\ref{eq:csym_fluct_fr}) with bulk symmetry energy (dashed line)
instead of fragments symmetry energy (solid line).
However it is important to observe that, for the light fragments which are isotopically
resolved in most experimental data-sets, $\sigma_I^2$
stays an excellent measure of the input symmetry energy irrespective
of conservation laws and Coulomb effects.

\begin{figure}
\resizebox{0.8\textwidth}{!}{%
  \includegraphics{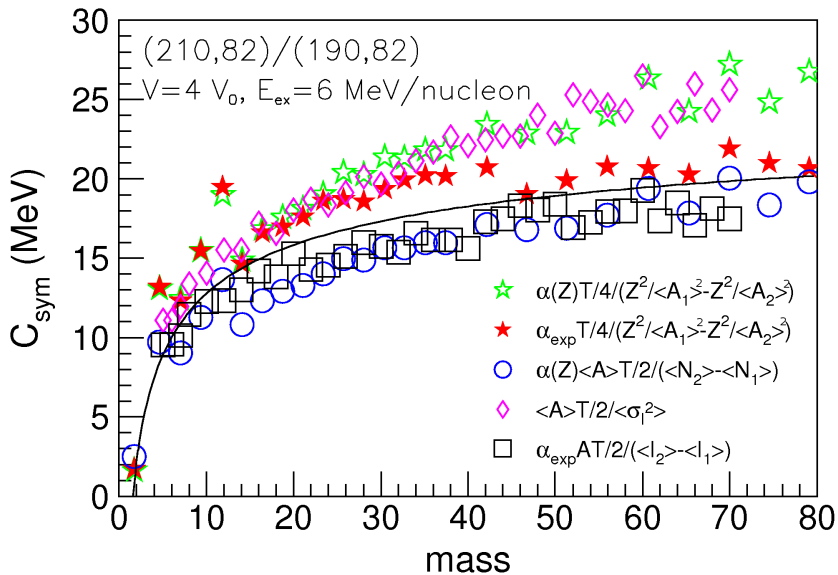}
  }
  \caption{
(Color online)
    The same as in Fig. \ref{fig:casym_noc}, but for the break-up stage of 
    (210,82) and (190,82) nuclei with $V =4V_0$ and $E_{ex}$=6 MeV/nucleon 
    when the Coulomb energy is included. The estimation of Eq. (\ref{eq:isoscaling_flucti})
    (open squares) is also presented. For this last expression, the abscissa has the meaning 
    of the exact mass.
  }
\label{fig:casym+c}
\end{figure}

Finally, Fig. \ref{fig:casym+c} gives the quality of the approximations
of the different formulas (Eqs. (\ref{eq:csym_fluct_fr}), (\ref{eq:isoscaling_fluct}), 
(\ref{eq:isoscaling_flucti}) and (\ref{eq:csym_ono}))
proposed in the previous section to access the symmetry energy.
As in the previous section, for the so-far widely used Eq. (\ref{eq:csym_ono})
we considered two 'definitions' of $\alpha$:
the slope of the logarithm of isotopic yield ratios for each $Z$ 
(Eq. (\ref{eq:isoscaling})) and, in the spirit of most experimental analyses,
the average value of $\alpha(Z)$ for fragments with $1 \leq Z \leq 8$. 
Not surprisingly, the reconstruction of the input symmetry energy is
perturbed by Coulomb effects.
As one may notice, for heavy fragments,
the violation of the different approximations
on which the proposed expressions reckon 
leads to a relative dispersion which may range
from 15\% (for $A=20$) to 25\% (for $A=70$).
The best description of the fragment symmetry energy seems to be offered by the
simplified version of Eq. (\ref{eq:isoscaling_flucti}) where 
the parameter $\alpha$ is calculated as the average
value of Eq. (\ref{eq:isoscaling}) over fragments with $Z$=1 to 8
as it underestimates the witness value by less than 1 MeV
over the whole considered fragment mass interval ($A=1-70$).
As a general statement, considering the light ($A\leq 20$) fragments
used in most experimental analyses for which the conservation laws constraint
are the least important, it is encouraging to 
see that both isoscaling and isotopic widths give an 
consistent estimation of $C_{sym}$ which deviates from the input value 
of no more than 20\% \cite{micro_casym}.

\section{Conclusions}

Isospin effects on the thermal and phase properties of finite uncharged 
excited nuclear systems at sub-nuclear densities have been studied 
in the framework of a microcanonical statistical model 
with cluster degrees of freedom.
We find that the isospin asymmetry of the source reduces the width of the
coexistence region and the critical temperature and pressure values,
in qualitative agreement with the well known results for infinite nuclear 
matter. The similarity of the phase diagram with the nuclear matter one
is an extra confirmation that the multifragment production phenomenon 
can be associated to the coexistence zone of a first order phase transition
of the liquid-gas type.

To push this connection further, we have explored the relation between 
fragment chemical composition and the expected 
isospin fractionation in the coexistence region of a multi-fluid system.
We have shown that a number of excess neutrons are emitted as free
particles in neutron-rich sources. This number strongly increases with 
the asymmetry of the source, independent of the system location in 
the phase diagram.
This implies that free nucleons emitted by a finite isolated system
cannot be unambiguously associated to the gas phase of the corresponding
phase diagram. Because of the mass and charge conservation law, the 
isotopic distribution of complex fragments is in turn mainly dictated by the 
excess free nucleons. Thus the fractionation phenomenon cannot be 
taken as a measure of phase coexistence. 
An interesting observable is given by the 'U-shape' of the
complete $Z/A$ vs. $Z$ distribution which appears characteristic of
the phase coexistence region, 
while more simple quantities like ratios of mirror nuclei
give ambiguous results.

Finally, we investigate the relation between fragments symmetry energy and the
variance of isotopic distributions. A simple expression relates
the symmetry energy with the isospin asymmetry variance 
and a new formula is proposed in order to calculate the
same quantity out of isoscaling observables.
While they are always accurate enough in the case of finite neutral systems,
some of these expressions show sizable deviations in real nuclei under
the influence of Coulomb. The present study, however, suggests that the widths of
the isotopic distributions, as well as the isoscaling parameter,
can still give a correct estimation of the symmetry energy in 
physical nuclear multifragmentation data, provided the break-up fragment partitions
can be restored from the detected cold fragments.

\end{document}